\documentclass[twocolumn,aps,prb,amsmath,amssymb,superscriptaddress]{revtex4-2}

\usepackage{newtxtext}
\usepackage{newtxmath}
\usepackage{microtype}
\usepackage{wasysym}
\usepackage{graphicx}
\usepackage{dcolumn}
\usepackage{bm}
\usepackage{multirow}
\usepackage{amssymb}
\usepackage{graphicx}
\usepackage{xcolor}
\definecolor{TighnariBrown}{RGB}{141,87,41}
\definecolor{TighnariGreen}{RGB}{40,114,70}
\definecolor{TighnariYellow}{RGB}{247,191,99}
\definecolor{PRLBlue}{RGB}{46,48,146}
\usepackage[
    colorlinks,
    citecolor=PRLBlue,
    linkcolor=PRLBlue,
    urlcolor=PRLBlue,
]{hyperref}
\usepackage{indentfirst}
\usepackage{cases}

\begin{document}
\title{Universal Short-Imaginary-Time Quantum Critical Dynamics Near Boundaries}
\author{Yu-Rong Shu}
\email{yrshu@gzhu.edu.cn}
\affiliation{School of Physics and Materials Science, Guangzhou University, Guangzhou 510006, China}

\author{Yuan-Biao Li}
\affiliation{School of Physics and Materials Science, Guangzhou University, Guangzhou 510006, China}

\author{Shuai Yin}
\email{yinsh6@mail.sysu.edu.cn}
\affiliation{School of Physics, Sun Yat-sen University, Guangzhou 510275, China}
\affiliation{Guangdong Provincial Key Laboratory of Magnetoelectric Physics and Devices, Sun Yat-sen University, Guangzhou 510275, China}

\date{\today}

\begin{abstract}

  While imaginary-time evolution has long served as a standard paradigm for ground-state preparation in numerical simulations and quantum devices, its intrinsic dynamical properties has been largely overlooked.
  Here, we investigate the short-imaginary-time critical dynamics in quantum systems with boundaries. A universal scaling theory is developed and verified in the two-dimensional quantum Ising model, uncovering rich dynamic critical behaviors dictated by boundary universality classes.
  For ordered initial states, the boundary order parameter $M_s$ decays with imaginary time $\tau$ as $M_s \propto \tau^{-\beta_1/\nu z}$, where $\beta_1$ denotes the boundary order parameter exponent, and $\nu$ and $z$ correspond to the correlation length exponent and the dynamic exponent, respectively.
  For disordered initial states, the autocorrelation of the boundary order parameter is governed by a novel critical exponent $\theta_1$, which is closely related to the critical initial slip behavior of $M_s$ characterized by the corresponding exponent $\theta_1'$.
  In contrast to its positive bulk counterpart, the boundary initial-slip exponent $\theta_1'$ is negative for the ordinary transition while remaining positive for the special transition.
  Although the static universality classes of $d$-dimensional quantum phase transitions generally coincide with those of $(d+1)$-dimensional classical phase transitions, we show that $\theta_1$ does not follow this conventional quantum-classical mapping.
  We further discuss the implications of our results for more exotic forms of boundary criticality.
  Our findings provide new physical insights into boundary critical dynamics and offer a novel route for probing exotic boundary critical behaviors in quantum many-body systems.
\end{abstract}

\maketitle

%%%%%%%%%%%%%%%%%%%%%%%%%%%%%%%%%%%%%%%%%%%%%%%%%%%%
{\bf Introduction}---Boundaries pervade realistic material systems. Boundary criticality constitutes a long-standing core problem across statistical and condensed-matter physics, since interfacial boundaries break the bulk translational invariance and substantially modify the critical scaling behaviors associated with phase transitions~\cite{PTCP8,PTCP10,Diehl1997book,Pleimling2004jpa,Binder1972prb,Binder1984prl,Landau1990prb,Ruge1995prb,Deng2005pre,Deng2006pre,Ritschel1995prl,Pleimling2005prb,Pleimling2004prl,Lin2008pre}.
While the critical properties of bulk systems are well established and uniquely determined by dimensionality, intrinsic symmetry and relevant perturbations, boundaries introduce additional interfacial degrees of freedom and give rise to critical behaviors distinct from the bulk critical phenomena.
Recently, the discovery of the extraordinary-log universality class~\cite{Metlitski2022sp} and related unconventional boundary critical behaviors~\cite{Toldin2021prl,Hu2021prl,Toldin2022prl,Padayasi2022sp,Sun2022prb,Zhang2022prb,Sun2022prb_log,Sun2025sp,Hu2026prl}, together with increasing attention to boundary effects in quantum critical systems, has renewed broad interest in this field~\cite{Zhang2017prl,Zhu2025arx,Wang2024prb,Liu2024prl,Liu2025arx,Toldin2025arx,Wu2020prb,Scaffidi2017prx,Ding2018prl,Weber2018prb,Verresen2021prx,Zhu2021prb,Yu2022prl}.

Beyond equilibrium critical behavior, critical systems exhibit universal short-time dynamics starting from prescribed initial states, which has become a fundamental topic in the study of classical phase transitions. One of the most remarkable features of short-time critical dynamics is the critical initial slip, which requires an independent exponent $\theta$ apart from conventional equilibrium critical exponents~\cite{Janssen1989zpb}. In classical systems, this exponent is governed by the equilibrium critical fixed point and provides a universal characterization of the early-time evolution~\cite{Janssen1989zpb,Li1995prl,Zheng1996prl,Albano2011rpp}.

Short-time critical dynamics also emerges in quantum phase transitions.
Real-time evolution after a quantum quench can also exhibit universal short-time scaling in a prethermal regime~\cite{Chiocchetta2015prb,Chiocchetta2016prb,Chiocchetta2017prl,Jian2019prl,Yin2021prb} . However, due to the unitary nature of the real-time evolution, this scaling is generally associated with a dynamical critical point, rather than being directly controlled by the ground-state quantum critical point.

In contrast, imaginary-time critical dynamics is directly controlled by the underlying ground-state quantum critical point, providing a unique setting in which nonequilibrium relaxation and equilibrium criticality become closely connected~\cite{Yin2014prb,Zhang2014pre}.
As a widely adopted technique for preparing many-body ground states in numerical simulations~\cite{Sandvik2010review,Vidal2004prl,Vidal2007prl,Jordan2008prl,Jiang2008prl,Assaad2008book,Li2019review} and on quantum computing platforms~\cite{Motta2020natphy,Nishi2021npj,Huo2023quantum}, imaginary-time evolution has traditionally been employed to access equilibrium ground-state properties, while the relaxation process itself contains rich nonequilibrium information.
For a fully ordered initial state, it was shown that the order parameter evolves with the imaginary time $\tau$ as $M\propto\tau^{-\beta/\nu z}$ at the critical point, where $\beta$ is the order parameter exponent, $\nu$ is the correlation length exponent and $z$ is the dynamic exponent.
Instead, starting from a disordered initial state with a small initial magnetization $m_0$, $M$ evolves according to $M\propto m_0\tau^\theta$, in which $\theta$ is the critical initial-slip exponent~\cite{Yin2014prb,Zhang2014pre}.
These scaling relations enable the determination of critical properties directly from the early stage of imaginary-time evolution without projecting the system to its ground state.
This feature substantially reduces the computational cost and has been successfully applied to a broad range of quantum critical systems, including correlated fermionic systems affected by the sign problem and quantum circuits~\cite{Yin2014prb,Zhang2014pre,Shu2017prb,Shu2022prl,Shu2022prb,Yu2026prl,Zhang2024prb,Lin2026prb,Yu2026sa,Shen2025arx}. 

While boundary critical relaxation has been extensively investigated in classical systems~\cite{Dietrich1983zpb,Kikuchi1985prl,Diehl1994prb,Ritschel1995prl,Pleimling2004prl,Pleimling2005prb,Lin2008pre} and short-imaginary-time critical dynamics has been established for quantum phase transitions~\cite{Yin2014prb,Zhang2014pre,Shu2017prb,Shu2022prl,Yu2026prl}, the imaginary-time relaxation dynamics near boundaries of quantum systems remain largely unexplored.

In this paper, we systematically investigate the short-imaginary-time critical dynamics near the boundaries of quantum systems. Taking the two-dimensional ($2$D) quantum Ising model as a prototypical example, we study the ordinary and special boundary universality classes starting from
different initial states. For the ordered initial states, we find that the boundary order parameter $M_s$ obeys the scaling relation $M_s\propto \tau^{-\beta_1/\nu z}$, where $\beta_1$ is the boundary order parameter exponent. For disordered initial states, we show that an additional critical exponent $\theta_1$ (or $\theta_1'$) should be introduced to characterize the critical initial slip behaviors at the boundary.
Remarkably, $\theta_1$ ($\theta_1'$) is negative for the ordinary transition but positive for the special transition.
We further uncover the scaling relation between $\theta_1$ ($\theta_1'$) and the bulk initial-slip exponent $\theta$.
These results establish the framework of short-imaginary-time critical dynamics near boundaries and reveal rich nonequilibrium scaling behavior associated with boundary universality classes, thereby bringing boundary criticality into the realm of short-imaginary-time dynamics.

{\bf Scaling theory}---In equilibrium, the bulk and boundary order parameters scale as $M_b\propto |g|^\beta$ and $M_s\propto |g|^{\beta_1}$, respectively, where $\beta$ and $\beta_1$ are the bulk and boundary order parameter exponents, and $g$ is the distance to the critical point. Near the boundary, the distance from the boundary $x$ introduces an additional length scale, and the local order parameter satisfies the scaling form $M_s(x,g)=|g|^\beta f_s(x|g|^\nu)$~\cite{PTCP8,PTCP10,Diehl1997book}.
Accordingly, to recover the boundary scaling behavior $M_s\propto |g|^{\beta_1}$, when $x\ll \xi$ (with $\xi$ being the correlation length), $M_s$ must satisfy $M_s\propto x^{(\beta_1-\beta)/\nu}|g|^{\beta}|g|^{\beta_1-\beta}$, yielding the short-distance expansion (SDE) of $M_s$~\cite{Symanzik1981npb,PTCP8,PTCP10,Diehl1997book,Ritschel1995prl}.

The SDE has been successfully applied to classical boundary critical dynamics~\cite{Ritschel1995prl,Pleimling2005prb,Pleimling2004prl,Lin2008pre}. Here, we generalize the SDE to quantum short-imaginary-time critical dynamics and establish a general scaling theory for universal critical dynamics near boundaries under different initial conditions.
From an ordered initial state, the bulk order parameter $M_b$ follows $M_b=\tau^{-\beta/\nu z}f_b(\tau L^{-z})$ at the critical point~\cite{Yin2014prb,Zhang2014pre}. The boundary order parameter $M_s$ is expected to obey the scaling form
$M_s(\tau,x,L)=\tau^{-\beta/\nu z}f_1(\tau x^{-z},\tau L^{-z})$.
When $\tau\ll L^z$, finite-size effects can be neglected.
Extending the SDE to the dynamic case, consistency with the boundary scaling behavior requires
$f_1\propto x^{(\beta_1-\beta)/\nu} \tau^{-(\beta_1-\beta)/\nu z}$.
Therefore, in the short-time regime, $M_s$ behaves as
\begin{equation}
M_s(\tau)\propto \tau^{-\beta_1/\nu z},
\label{eq:ms}
\end{equation}
with the scaling form
\begin{equation}
M_s(\tau,L)=\tau^{-\beta_1/\nu z}f_2(\tau L^{-z}).
\label{eq:ms1}
\end{equation}

From an uncorrelated disordered initial state, we first consider the equal-time spatial correlation, defined as $C(r,\tau)\equiv \langle \sigma^z_0(\tau)\sigma^z_r(\tau) \rangle$, with $r$ being the spatial distance. Since the correlation length $\xi$ grows as $\xi\propto \tau^{1/z}$, in the bulk, the correlation function is expected to increase as $C\propto r^{-(1+\eta)}\exp(-cr/\tau^{1/z})$, where $\eta$ is the anomalous dimension and $c$ is a constant. When both operators 
are located along the boundary direction, applying the SDE to both fields yields the longitudinal correlation function, which satisfies
\begin{equation}
C_{\parallel}(r,\tau)\propto r^{-(1+\eta_{\parallel})}\exp(-a r/\tau^{1/z}),
\label{eq:para}
\end{equation}
in which $a$ is constant, ${\parallel}$ denotes the direction along the boundary and $r$ measures the distance in this direction. The exponent is given by $\eta_{\parallel}=\eta+2(\beta_1-\beta)/\nu$~\cite{Diehl1997book}.
For a boundary-bulk pair of operators, the SDE acts only on the boundary field, leading to the transverse correlation
\begin{equation}
C_{\perp}(r,\tau)\propto r^{-(1+\eta_{\perp})}\exp(-br/\tau^{1/z}).
\label{eq:perp}
\end{equation}
Similar to $C_{\parallel}$, here $b$ is a prefactor and $r$ is the distance along the direction perpendicular to the boundary denoted by ${\perp}$. The exponent is given by $\eta_{\perp}=(\eta+\eta_\parallel)/2$~\cite{Diehl1997book}.

Apart from the buildup of spatial correlations, a distinct feature of relaxation from a disordered initial state is the critical initial slip. This behavior is reflected in the autocorrelation function, defined as $A(\tau)\equiv \langle \sigma^z_r(0)\sigma^z_r(\tau) \rangle$. In the bulk, the autocorrelation $A_b$ obeys $A_b\propto \tau^{\theta-d/z}$~\cite{Yin2014prb,Yu2026prl}.
At the boundary, since the SDE acts on both $\sigma_r^z$-operators in $A_s$, one arrives at
\begin{equation}
A_s(\tau,L)=\tau^{\theta_1-d/z}f_3(\tau L^{-z}),
\label{eq:autoc}
\end{equation}
with the exponent
\begin{equation}
\theta_1=\theta-2(\beta_1-\beta)/\nu z,
\label{eq:scarela}
\end{equation}
being the critical initial slip exponent at the boundary.

The critical initial slip can also be characterized by the short-imaginary-time evolution of the order parameter in the presence of a small initial magnetization $m_0$~\cite{Yin2014prb}. Near the boundary, $M_s(x,\tau)=m_0\tau^{\theta}f_4(x \tau^{-1/z})$. Applying the SDE to the order-parameter operator yields 
\begin{equation}
M_s(\tau)\propto m_0\tau^{\theta_1'}.
\label{eq:msslip}
\end{equation}
in which
\begin{equation}
\theta_1'=\theta-(\beta_1-\beta)/\nu z.
\label{eq:theta1p}
\end{equation}
Comparing the exponents $\theta_1$ and $\theta_1'$, one finds that in the short-time regime, both $M_s$ and $A_s$ are governed by related exponents, with the difference arising from whether the SDE acts on one or two order-parameter operators.

The above analysis can be readily generalized to full scaling forms that incorporate the distance from the boundary, thereby describing the crossover from bulk to boundary critical dynamics. Details are presented in Sec.~I of the Supplemental Materials.

{\bf Model and numerical method}---To verify the scaling theory, we consider the $2$D quantum Ising model on a square lattice, described by the Hamiltonian
\begin{equation}
H = -J\sum_{\langle i,j \rangle} \sigma_i^z \sigma_j^z- h\sum_{i} \sigma_i^x-J_s\sum_{\langle i,j \rangle'} \sigma_i^z \sigma_j^z ,
\label{eq:ham}
\end{equation}
where $\sigma_i^{x,z}$ are Pauli matrices, $h$ is the strength of the transverse field, and $J_s$ is the coupling strength of boundary bonds (denoted by $\langle i,j\rangle'$), while all remaining bonds (denoted by $\langle i,j\rangle$) have coupling strength $J$. 
In the following, we set $J=1$. To investigate boundary effects, we impose periodic boundary conditions along the $x$ direction and open boundary conditions along the $y$ direction, such that the boundaries extend along the $x$ direction. The lattice size is chosen as $2L\times L$.

The equilibrium critical properties of this model are well-established. The bulk quantum critical point is located at $h_\mathrm{c}=3.04451(7)$~\cite{Shu2017prb}, and is characterized by the critical exponents $\beta=0.32653(10)$~\cite{Campostrini2002pre}, $\nu=0.6299709(40)$~\cite{Simmons-Duffin2017} and $z=1$~\cite{Sachdev1999book}.
Depending on the boundary coupling strength $J_s$, different boundary universality classes can emerge. For weak boundary couplings, the boundary orders simultaneously with the bulk at the bulk critical point, corresponding to the ordinary universality class with boundary order-parameter exponent $\beta_1=0.8003(4)$~\cite{Hasenbusch2011prb_2}.
For sufficiently strong boundary couplings, the boundary is already ordered at the bulk critical point, giving rise to the extraordinary universality class.
At the critical boundary coupling $J_{s}^{*}=1.9895(6)$, the boundary critical behavior belongs to the special universality class, which separates the ordinary and extraordinary regimes, with boundary exponent $\beta_1=0.2227$~\cite{Hasenbusch2011prb_3}. For the determination of $J_{s}^{*}$, see Sec.~II of the Supplemental Materials.

To investigate the short-time critical dynamics, we use the projector quantum Monte Carlo (QMC) method to simulate the imaginary-time evolution $|\psi(\tau)\rangle=e^{-\tau H}|\psi(0)\rangle$~\cite{Sandvik2010review,Sandvik2003pre,Grandi2011prb,Grandi2013jpcm,Liu2013prb}.
The imaginary-time evolution operator $e^{-\tau H}$ is series-expanded into a sequence of operators acting on the basis states.
Efficient importance-sampling procedures of imaginary-time evolution, including local single-bond and global cluster updates, are carried out. The computation cost of a full Monte Carlo sweep scales as $\tau L^{d}$.
In QMC simulations aiming at the short-time critical dynamics, the preparation of the initial state is crucial.
The initial condition is encoded in the space-time sampling configuration at $\tau=0$. For the ordered initial state, the space-time sampling boundary is fixed to a fully ordered state, whereas for the disordered initial state, it is sampled from uncorrelated spin configurations with $\langle m_0\rangle=0$. For a more detailed introduction of the method, we refer to the literature~\cite{Sandvik2010review,Sandvik2003pre,Grandi2011prb,Grandi2013jpcm,Liu2013prb}.

\begin{figure}[tbp]
\centering
  \includegraphics[width=1\linewidth,clip]{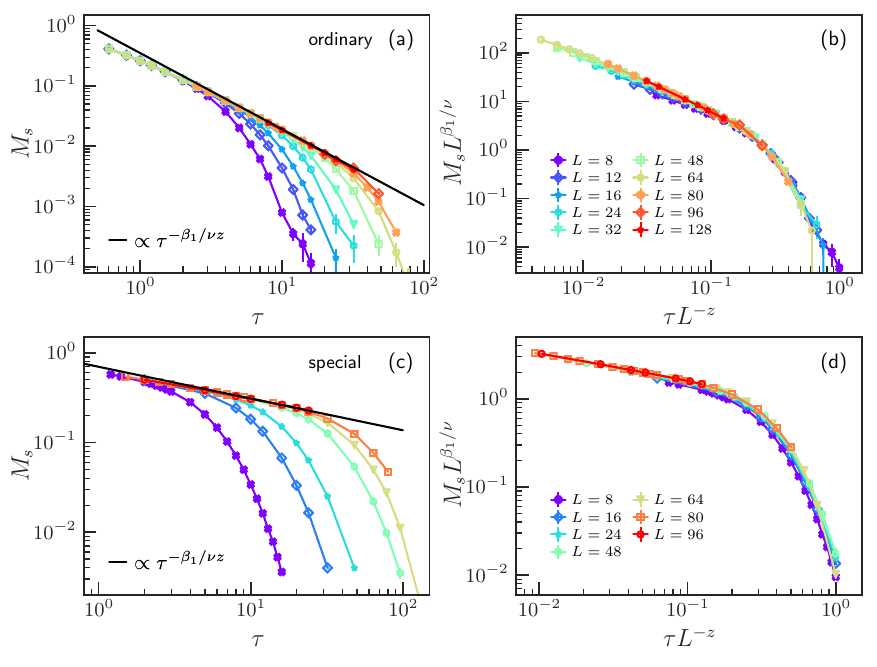}
  \vskip-3mm
  \caption{{\bf Dynamics of the boundary order parameter $M_s$ from the ordered initial state.}
    (a) and (b): Evolution of $M$ before and after rescaling for the ordinary transition.
    (c) and (d): Evolution of $M$ before and after rescaling for the special transition.
The solid lines indicate power-law decay with the exponent $\beta_1/\nu z=1.2704$ and $0.3535$ for the ordinary and special transition, respectively. Log-log scale is used in all plots.
   }
  \label{fig1}
\end{figure}

{\bf Ordered initial state}---At the ordinary transition, the boundary order parameter $M_s$ exhibits a clear power-law decay after a microscopic transient regime. As shown in Fig.~\ref{fig1}(a), the data for different system sizes follow the scaling behavior $M_s\propto \tau^{-\beta_1/\nu z}$ over a broad temporal window when $\tau \ll L^{z}$, indicating that the imaginary-time relaxation at the boundary is governed by $\beta_1$, in agreement with Eq.(\ref{eq:ms}). When $\tau\sim L^{z}$, finite-size effects drive the curves to deviate from the asymptotic scaling regime. After rescaling $M_s$ and $\tau$ as $M_sL^{\beta/\nu}$ and $\tau L^{-z}$, respectively, we find that the rescaled curves collapse well, as shown in Fig.~~\ref{fig1}(b), demonstrating that $M_s$ obeys the full scaling form $M_s=L^{-\beta_1/\nu}g(\tau L^{-z})$, which is equivalent to Eq.~(\ref{eq:ms1}) after identifying $g(\tau L^{-z})=(\tau L^{-z})^{-\beta_1/\nu z} f_1(\tau L^{-z})$.

Similar scaling behaviors are observed in the special transition. As shown in Fig.~\ref{fig1}(c), in the universal short-time stage, $M_s$ decays as $M_s\propto \tau^{-\beta_1/\nu z}$ with $\beta_1=0.2227(4)$~\cite{Hasenbusch2011prb_3} for the special transition. The good collapse of $M_sL^{\beta_1/\nu}$ and $\tau L^{-z}$ for different system sizes again verifies the scaling form in Eq.~(\ref{eq:ms1}).

\begin{figure}[tbp]
\centering
  \includegraphics[width=1\linewidth,clip]{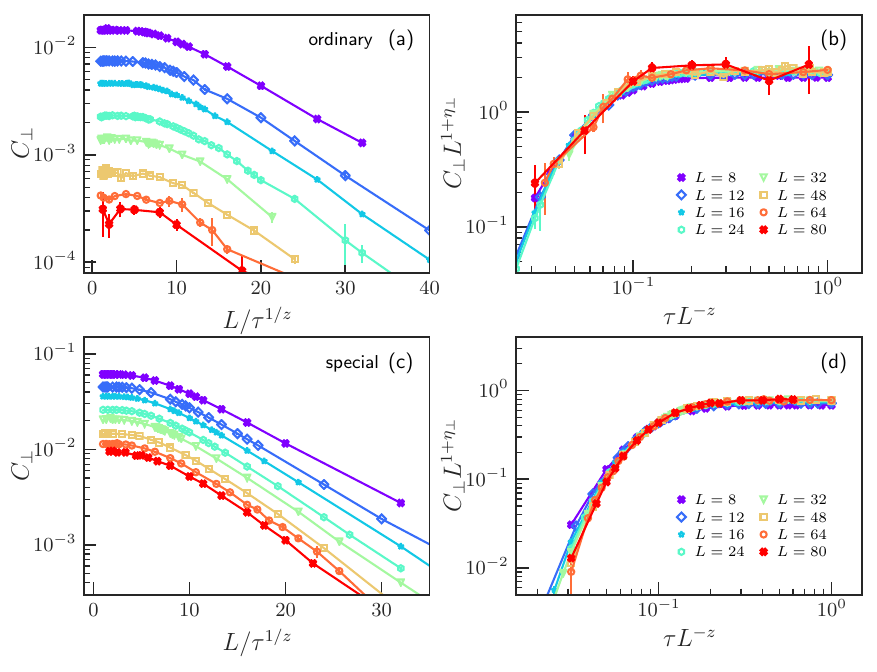}
  \vskip-3mm
  \caption{{\bf Dynamics of the transverse correlation $C_{\perp}$ from the disordered initial state.}
    (a): Dependence of $C_{\perp}$ on $L/\tau^{1/z}$ for the ordinary transition with $\eta_{\perp}=1.7887$. 
    (b): Finite-size rescaling collapse of the data in (a).
    (c): Same as (a), but for the special transition, $\eta_{\perp}=0.8718$.
    (d): Finite-size rescaling collapse of the data in (c).
    Semi-log scale is used in (a), (c), while (b), (d) use log-log scale. The straight lines in (a) and (c) indicate exponential dependence.
   }
  \label{fig2}
\end{figure}

{\bf Disordered initial state}---The short-imaginary-time dynamics starting from a disordered initial state exhibits qualitatively different behavior. 
Figures~\ref{fig2}(a) and (c) show the dependence of $C_{\perp}(L/2,\tau)$ on $L/\tau^{1/z}$ for the ordinary and special universality classes, respectively, in a semi-log scale. In the short-time regime (large $L/\tau^{1/z}$), $C_{\perp}(L/2,\tau)$ follows the exponential form $\exp(-aL/\tau)$, consistent with Eq.~(\ref{eq:perp}). 
In the semi-log scale, the data for different system sizes fall on straight lines that are nearly parallel, indicating that the coefficient $a$ is independent of $L$. 
As $\tau$ increases, $C_{\perp}(L/2,\tau)$ gradually approaches it equilibrium value.
After rescaling $\tau$ and $C_\perp$ as $\tau L^{-z}$ and $C_\perp L^{1+\eta_\perp}$, respectively, the rescaled curves collapse well for both cases, confirming Eq.~(\ref{eq:perp}), as shown in Figs.~\ref{fig2}(b) and (d). The value of $\eta_{\perp}$ is $1.7887$ and $0.8718$, for the ordinary and special universality class, respectively~\cite{Campostrini2002pre,Simmons-Duffin2017,Hasenbusch2011prb_2,Hasenbusch2011prb_3}.

The longitudinal correlation exhibits a similar exponential dependence $C_{\parallel}\propto\exp(-bL/\tau^{1/z})$ on $L/\tau^{1/z}$ in the short-time regime for both the ordinary and special universality classes, as shown in Figs.\ref{fig3}(a) and (c), respectively. In the semi-log scale, the slopes of the lines are close, further suggesting that the coefficient $b$ is independent of $L$.
In the long-time (small $L/\tau^{1/z}$) regime, $C_{\parallel}(L/2,\tau)$ approaches its equilibrium value. In Figs.\ref{fig3}(b) and (d), one finds that the rescaled curves of $C_\parallel L^{1+\eta_\parallel}$ versus $\tau L^{-z}$ collapse well, with the exponent $\eta_\parallel=2.5408$ for the ordinary transition and $\eta_{\parallel}=0.7070$ for the especial transition~\cite{Campostrini2002pre,Simmons-Duffin2017,Hasenbusch2011prb_2,Hasenbusch2011prb_3}, confirming the validity of Eq.~(\ref{eq:para}). 

%A notable difference between the two boundary universality classes is the magnitude of the correlations. For both $C_\perp$ and $C_\parallel$, the correlation amplitudes at the special transition are significantly larger than those at the ordinary transition throughout the imaginary-time evolution, reflecting the stronger boundary critical correlations associated with the special universality class.

\begin{figure}[tbp]
\centering
  \includegraphics[width=1\linewidth,clip]{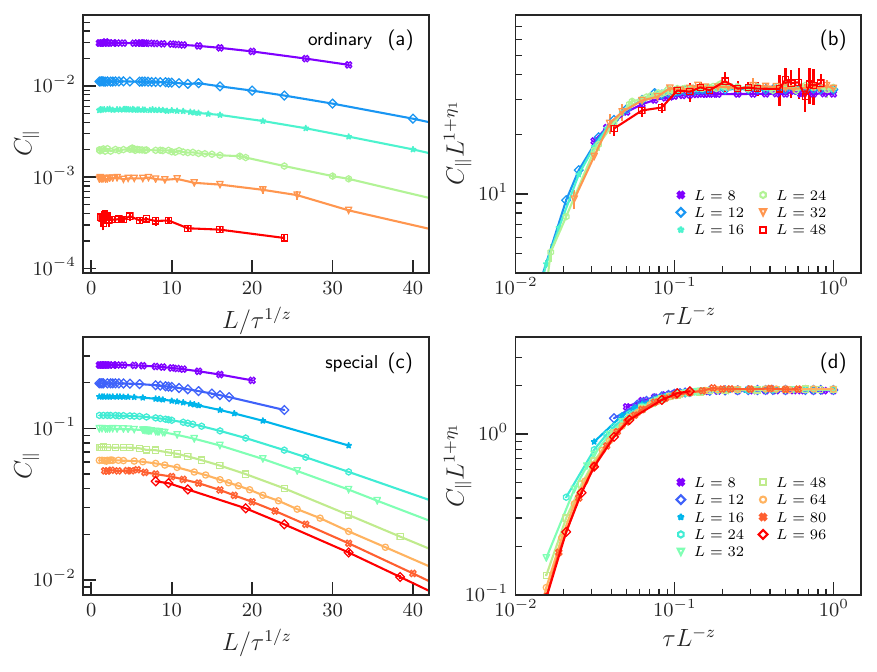}
  \vskip-3mm
  \caption{{\bf Dynamics of the longitudinal correlation $C_{\parallel}$ near the boundary from the disordered initial state.}
    (a): Dependence of $C_{\parallel}$ on $L/\tau^{1/z}$ for the ordinary transition with $\eta_{\parallel}=2.5408$.
    (b): Finite-size rescaling collapse of the data in (a).
    (c): Same as (a), but for the special transition $\eta_{\parallel}=0.7070$.
    (d): Finite-size rescaling collapse of the data in (c).
    Semi-log scale is used in (a), (c), while (b), (d) use log-log scale. The straight lines in (a) and (c) indicate exponential dependence.
   }
  \label{fig3}
\end{figure}

{\bf Critical initial slip near the boundary}---The imaginary-time relaxation from a disordered initial state also captures the hallmark feature of short-time critical dynamics, namely the critical initial slip characterized by the exponent $\theta$. Near the boundary, the SDE predicts that this initial-slip behavior is instead governed by a distinct exponent, $\theta_1$, which is manifested in the boundary autocorrelation function.

To verify this prediction, we compute the boundary autocorrelation functions $A_s$ for the ordinary and special universality classes, as shown in Figs.\ref{fig4}(a) and (c), respectively. For both cases, we find that $A_s$ exhibits a power-law decay $A_s\propto\tau^{\theta_1-d/z}$ in the short-imaginary-time regime. The value of $\theta_1$ is determined by Eq.~(\ref{eq:scarela}), giving $\theta_1\approx -1.295$ for the ordinary transition, and $\theta_1\approx 0.539$ for the special transition.
After rescaling $A_s$ and $\tau$ as $A_sL^{d-\theta_1 z}$ and $\tau L^{-z}$, we find that the rescaled curves collapse well, confirming that Eq.~(\ref{eq:autoc}) holds for both the ordinary and special transitions, as shown in Figs.\ref{fig4}(b) and (d).

\begin{figure}[tbp]
\centering
  \includegraphics[width=1\linewidth,clip]{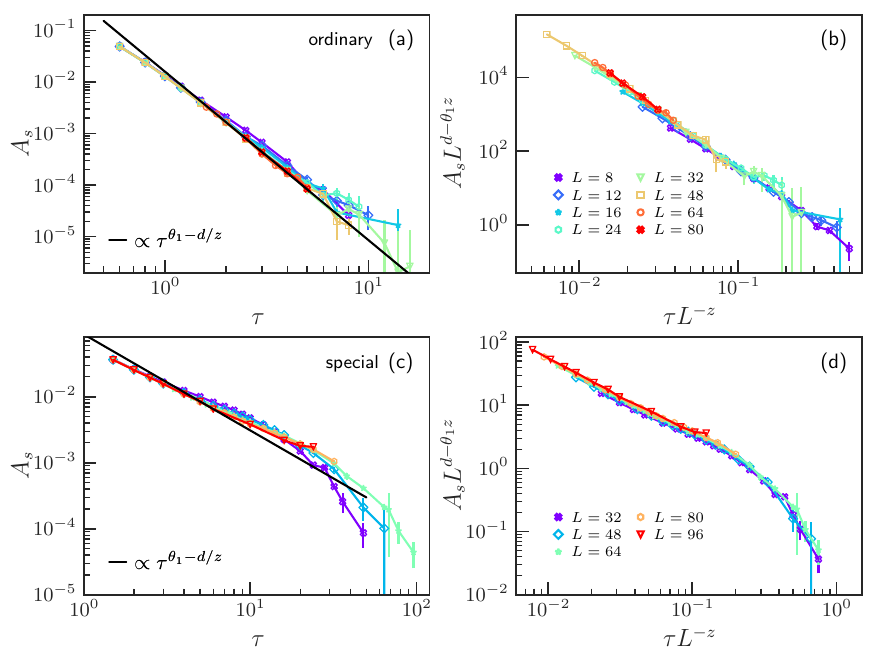}
  \vskip-3mm
  \caption{{\bf Dynamics of the boundary autocorrelation $A_s$ from the disordered initial state.}
    (a): Imaginary-time evolution of $A_s$ for the ordinary transition.
    (b): Finite-size rescaling collapse of the data in (a).
    (c): Same as (a), but for the special transition.
    (d): Finite-size rescaling collapse of the data in (c).
    The solid lines indicate power-law decay with the exponent $\theta_1-d/z$, with $\theta_1=-1.295$ and $0.539$ for the ordinary and special transition, respectively. Log-log scale is used in all plots.
   }
  \label{fig4}
\end{figure}

Furthermore, based on the numerical results presented above, we observe that the scaling form governing boundary criticality in short-imaginary-time dynamics coincides with that of short-time critical dynamics in classical systems.
However, unlike the static exponent $\beta_1$, which obeys the quantum-classical correspondence~\cite{PTCP8,PTCP10,Diehl1997book}, the boundary initial-slip exponent $\theta_1$ is an intrinsic dynamical exponent and therefore generally takes distinct values in the $2$D quantum and $3$D classical systems.
The corresponding values for the $3$D classical Ising universality class, according to Eq.~(\ref{eq:scarela}), are given by $\theta_1\approx -0.635$ and $0.271$ for the ordinary and special transitions, respectively~\cite{Ritschel1995prl}. The clear discrepancy from the quantum values indicates that the boundary initial-slip exponent does not follow the conventional quantum-to-classical correspondence.

%,Hasenbusch2020pre,Jaster1999jpa,Simmons-Duffin2017,Campostrini2002pre,Hasenbusch2011prb_2,Hasenbusch2011prb_3

%In addition, similar to $C_\perp$ and $C_\parallel$, $A_s$ is also larger at the special transition than at the ordinary transition. This enhancement originates from the stronger boundary couplings at the special transition, which lead to stronger temporal correlations at the boundary.

Moreover, the enhancement of the boundary coupling not only leads to quantitative differences in $A_s$ between the two boundary universality classes, but also qualitatively changes the boundary critical initial slip behavior. The boundary critical initial slip can also manifest itself in the short-imaginary-time evolution of the boundary order parameter $M_s$ with critical exponent $\theta_1'$ as shown in Eq.~(\ref{eq:msslip}). However, an accurate determination of $\theta_1'$ from Eq.~(\ref{eq:msslip}) requires an asymptotically small initial magnetization $m_0$~\cite{Yin2014prb}, which is difficult in finite-size simulations. Alternatively, given the validity of the SDE established above, $\theta_1'$ can be evaluated from Eq.~(\ref{eq:theta1p}) as $\theta_1'\approx -0.532$ for the ordinary transition and $\theta_1'\approx 0.376$ for the special transition, respectively. These values can also be verified by  a mixed two-time correlation, as presented in Sec.~III of the Supplemental Materials.
Unlike the bulk case, where the standard critical initial slip exponent $\theta$ is generally positive, $\theta_1'$ takes a negative value for the ordinary transition.

The opposite signs of $\theta_1'$ indicate qualitatively different boundary short-time dynamics in the two boundary universality classes. Physically, the critical initial slip arises from the competition between domain formation and critical fluctuations.
In the bulk, for a disordered state with small initial magnetization, ferromagnetic interactions tend to align neighboring spins, leading to the formation and growth of ferromagnetic domains, whereas critical fluctuations hinder this ordering process.
The positive bulk exponent $\theta\approx 0.209$~\cite{Shu2017prb} indicates that the growth of local order dominates in the early-time regime.

For the ordinary transition, the reduced coordination number at the boundary weakens the effective ferromagnetic coupling.
As a result, domain formation near the boundary is suppressed relative to the bulk, and critical fluctuations dominate the early boundary dynamics. This leads to a negative value of $\theta_1'$, rendering the order parameter decay in the short-time stage.

By contrast, at the special transition, the boundary coupling is sufficiently strong to induce a ferromagnetic phase transition at the boundary itself. Consequently, domain formation at the boundary is reinforced by the ordering process propagating from the bulk, leading to a stronger domain-growth mechanism and hence a positive $\theta_1'$. Furthermore, $\theta_1'$ is even larger than its bulk counterpart, indicating that boundary criticality further enhances domain growth.

{\bf Summary and discussion}---In this work, we investigate short-imaginary-time quantum critical dynamics near boundaries. We establish a scaling theory for boundary short-imaginary-time dynamics and identify a boundary initial-slip exponent $\theta_1$, which characterizes the influence of boundary criticality on universal early-time relaxation. The scaling theory is verified in the $2$D quantum Ising model for both the ordinary and special boundary universality classes.

A key finding is that the boundary critical initial-slip exponent takes opposite signs at the ordinary and special transitions. This qualitative difference originates from the competition between domain formation and critical fluctuations. For the ordinary transition, weakened boundary ordering suppresses domain formation, causing critical fluctuations to take control of the short-time dynamics and resulting in a negative $\theta_1'$. For the special transition, enhanced boundary correlations reinforce domain formation near the boundary and drive $\theta_1$ positive. Unlike the static boundary exponents, $\theta_1$ and $\theta_1'$ are dynamic exponents and generally differ between quantum and classical systems.

The present work establishes a general framework for boundary short-imaginary-time quantum critical dynamics.
The framework developed here naturally opens the door to studying more exotic forms of boundary criticality. One example is provided by topological boundaries~\cite{Zhang2017prl,Zhu2025arx,Wang2024prb,Liu2024prl,Liu2025arx,Toldin2025arx,Wu2020prb,Scaffidi2017prx,Ding2018prl,Weber2018prb,Verresen2021prx,Zhu2021prb,Yu2022prl}, where edge states may modify the propagation of critical correlations and lead to new dynamical scaling behavior.
Another intriguing case is the extraordinary-log universality class, where logarithmic boundary correlations replace conventional power-law scaling~\cite{Metlitski2022sp,Toldin2021prl,Hu2021prl,Hu2026prl,Toldin2022prl,Padayasi2022sp,Sun2022prb,Zhang2022prb,Sun2022prb_log,Sun2025sp}. How such unconventional boundary criticality manifests itself in short-imaginary-time relaxation and critical initial slip remains largely unexplored. Addressing these questions may uncover new dynamical signatures of boundary critical phenomena beyond the ordinary and special universality classes.

{\bf Acknowledgments}---S.Y. is supported by Quantum Science and Technology-National Science and Technology Major Project(Grant No.2025ZD0300400), the National Natural Science Foundation of China (Grants No. 12222515), the Research Center for Magnetoelectric Physics of Guangdong Province (Grant No. 2024B0303390001), the Guangdong Provincial Key Laboratory of Magnetoelectric Physics and Devices (Grant No. 2022B1212010008), the Science and Technology Projects in Guangzhou City (Grant No. 2025A04J5408).
Y.R.S. and Y.B.L. are supported by the National Natural Science Foundation of China (Grant No. 12104109), the Science and Technology Projects in Guangzhou (Grant No. 2024A04J2092).

%\textcolor{black}{\it Data availability}--- The data that support the findings of this article are openly available~[xxx].

\bibliographystyle{apsrev4-2}
\let\oldaddcontentsline\addcontentsline
\renewcommand{\addcontentsline}[3]{}
\bibliography{ref}
\let\addcontentsline\oldaddcontentsline
\onecolumngrid

\clearpage
\newpage
% \appendix
\widetext

\begin{center}
  \textbf{\large Supplemental Materials for ``Universal Short-Imaginary-Time Quantum Critical Dynamics Near Boundaries''}
\end{center}

\date{\today}
\maketitle

\renewcommand{\thefigure}{S\arabic{figure}}
\setcounter{figure}{0}
\renewcommand{\theequation}{S\arabic{equation}}
\setcounter{equation}{0}
\renewcommand{\thesection}{\Roman{section}}
\setcounter{section}{0}
\setcounter{secnumdepth}{4}

% \tableofcontents
% \hypersetup{linkcolor=blue}
\addtocontents{toc}{\protect\setcounter{tocdepth}{0}}
{
\tableofcontents
}
\section{Bulk-to-Boundary Scaling Crossover}

In the main text, we only discuss the scaling form of the local order parameter at the boundary. Here we consider the order parameter $M$ at location $x$. For the ordered initial state, the full scaling form is given by~\cite{Symanzik1981npb,PTCP8,PTCP10,Diehl1997book,Ritschel1995prl,Yin2014prb,Zhang2014pre}
\begin{equation}
  M(x,\tau,L)=\tau^{-\beta/\nu z}f_5(x\tau^{-1/z},\tau L^{-z}).
\label{eq:mxscaling}
\end{equation}
In the short-time regime $\tau\ll L^z$, finite-size effects can be neglected, yielding
\begin{equation}
  M(x,\tau)=\tau^{-\beta/\nu z}f_5(x\tau^{-1/z})
\label{eq:mxscaling1}
\end{equation}
with $f_5$ being the scaling function, satisfying
\begin{equation}
  f_5(u)\sim
  \begin{cases}
u^{(\beta_1-\beta)/\nu}, & u\ll 1,\\
{\rm const}, & u\gg1.
\end{cases}
\label{eq:f5}  
\end{equation}
The scaling variable $u=x\tau^{-1/z}=x/\xi$ compares the distance from the boundary with the growing correlation length $\xi\sim\tau^{1/z}$.
For the disordered initial state, the order parameter $M$ fluctuates around zero due to symmetry restriction. Thus, we consider the autocorrelation $A$ at different $x$ instead. The full scaling form of $A$ is given by~\cite{Symanzik1981npb,PTCP8,PTCP10,Diehl1997book,Ritschel1995prl,Yin2014prb,Yu2026prl}
\begin{equation}
  A(x,\tau,L)=\tau^{\theta-d/z}f_6(x\tau^{-1/z},\tau L^{-z}).
  \label{eq:ascaling}
\end{equation}  
Similarly, when $\tau\ll L^{z}$, the finite-size term can be dropped and one has
\begin{equation}
  A(x,\tau)=\tau^{\theta-d/z}f_6(x\tau^{-1/z}).
  \label{eq:ascaling1}
\end{equation}
In the bulk regime $u\gg 1$, the boundary is not yet probed and the bulk behavior is recovered; $f_6(u)$ tends to a constant.
In the boundary regime, the SDE acts twice and leads to 
\begin{equation}
A(x,\tau)\sim x^{2(\beta_1-\beta)/\nu}\tau^{-d/z+\theta_1},
\end{equation}
which requires $f_6(u)\sim u^{2(\beta_1-\beta)/\nu}$ when $u\ll 1$.
Therefore,
\begin{equation}
f_6(u)\sim
\begin{cases}
u^{2(\beta_1-\beta)/\nu},
&
u\ll 1,
\\{\rm const},
&
u\gg 1.
\end{cases}
\end{equation}

\begin{figure}[tbp]
\centering
  \includegraphics[width=0.75\linewidth,clip]{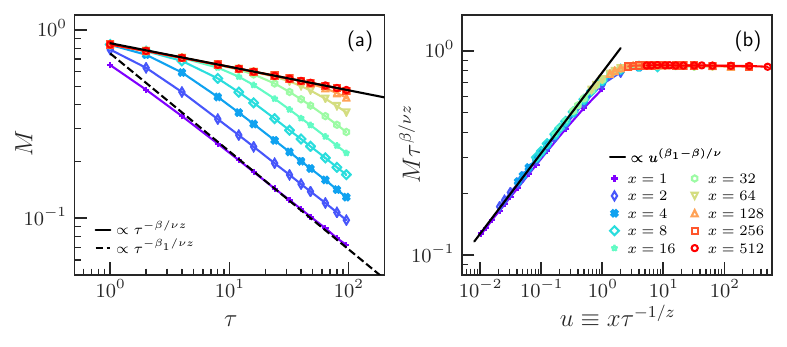}
  \vskip-3mm
  \caption{{\bf Scaling crossover of $M(x,\tau)$ for the $1$D quantum Ising model with the ordered initial state.} (a) Imaginary-time evolution of $M$ for different location $x$. The solid and dashed line represents power-law decay with the exponent $\beta/\nu z=1/8$ and $\beta_1/\nu z=1/2$, respectively. (b) Recaled curves of $M\tau^{\beta/\nu z}$ versus $x\tau^{-1/z}$. The solid line indicates power-law dependence of $[x\tau^{-1/z}]^{(\beta_1-\beta)/\nu}$.
   }
  \label{figs1}
\end{figure}

To verify the scaling forms derived above, we consider the $1$D quantum Ising model, for which the boundary critical exponents are known exactly.
Although in $1$D, the model contains only the ordinary boundary universality class, it provides a useful benchmark for testing the bulk-to-boundary crossover predicted by the SDE.
In the following, we examine both the local order parameter and autocorrelation function at different distances from the boundary.
To avoid finite-size effect, the system size is set as $L=1024$.

For the ordered initial state, Fig.~\ref{figs1}(a) shows the local order parameter $M(x,\tau)$ measured at different $x$ from the boundary. For sites far from the boundary, the decay follows the bulk critical behavior. As $x$ decreases, a clear crossover develops and the decay gradually approaches the boundary scaling behavior. This crossover reflects the growth of the correlation length during the imaginary-time evolution and the resulting influence of the boundary.
As shown in Fig.~\ref{figs1}(b), after rescaling all data collapse well onto a single curve, confirming the predicted crossover from bulk to boundary critical dynamics. 

\begin{figure}[tbp]
\centering
  \includegraphics[width=0.75\linewidth,clip]{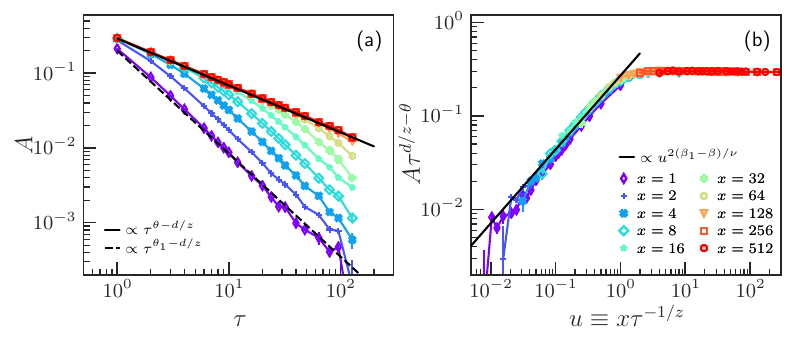}
  \vskip-3mm
  \caption{{\bf Scaling crossover of $A(x,\tau)$ for the $1$D quantum Ising model with the disordered initial state.}
    (a) Imaginary-time evolution of $A$ for different location $x$. The solid and dashed line represents power-law decay with the exponent $\theta-d/z=-0.6266$ and $\theta_1-d/z=-1.377$, respectively. (b) Recaled curves of $A\tau^{d/z-\theta}$ versus $x\tau^{-1/z}$. The solid line indicates power-law dependence of $[x\tau^{-1/z}]^{2(\beta_1-\beta)/\nu}$.
   }
  \label{figs2}
\end{figure}

For the disordered initial state, the $x$-dependence of the local autocorrelation function $A(x,\tau)$ is shown in Fig.~\ref{figs2}(a).
The decay of $A(x,\tau)$ exhibits a clear crossover from the bulk exponent $\theta-d/z$ to the boundary exponent $\theta_1-d/z$.
The good scaling collapse in Fig.~\ref{figs2}(b) strongly supports the proposed scaling form. The collapsed data exhibit the predicted asymptotic behavior $f_6(u)\sim u^{2(\beta_1-\beta)/\nu}$ for the boundary regime, while approaching a constant for the bulk regime, in excellent agreement with the scaling theory.

\section{Determination of the special transition point}

\begin{figure}[tbp]
\centering
  \includegraphics[width=0.5\linewidth,clip]{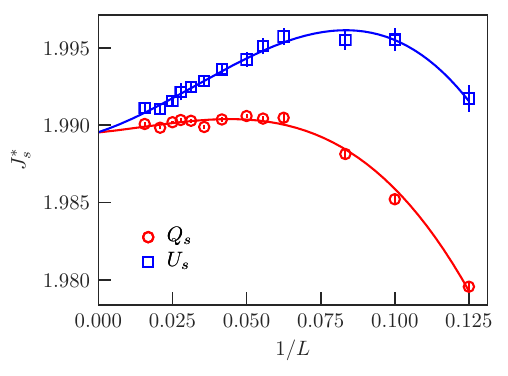}
  \vskip-3mm
\caption{{\bf Size-dependent crossing points of $Q_s$ and $U_s$}. The solid lines are fits to $J_s^\ast(L)=J_s^\ast+aL^{-\omega}$, yielding consistent $J_s^\ast=1.9895(6)$ in the thermodynamic limit.}
  \label{figqcL}
\end{figure}

The special transition point is determined from the imaginary-time relaxation starting from the fully ordered state. Following the procedure of Ref.~\cite{Shu2017prb}, we fix the ratio $\tau/L^z=1/4$ and compute two dimensionless quantities associated with the boundary order parameter.

The first quantity is the average sign of the boundary order parameter,
\begin{equation}
Q_s=\langle {\rm sgn}[M_s(\tau)]\rangle,
\end{equation}
which characterizes the memory of the initial ordered state during the relaxation. The second quantity is the Binder ratio
\begin{equation}
U_s=\frac{\langle M_s^2(\tau)\rangle}
{\langle M_s(\tau)\rangle^2}.
\end{equation}

At the special transition, both $Q_s$ and $U_s$ are scale invariant up to finite-size corrections. Consequently, curves for different system sizes intersect at $J_s^\ast$. Figure~\ref{figqcL} shows the finite-size crossing points of both quantities.
Extrapolating to the thermodynamic limit, we estimate
\begin{equation}
J_s^\ast=1.9895(6),
\end{equation}
which is used throughout this work.

\section{Critical initial slip near boundaries}

\subsection{Local two-time correlation}
A direct determination of the critical initial-slip exponent from the short-time evolution of the order parameter requires taking the limit $m_0\to 0$
~\cite{Yin2014prb,Zhang2014pre,Janssen1989zpb,Li1995prl,Zheng1996prl},
which is difficult to achieve in finite-size simulations. It is useful to consider the two-time correlation function instead, defined as
~\cite{Tome1998pre,Shu2017prb}
\begin{equation}
P(\tau)=L^{d-1}\langle M(\tau)M(0)\rangle,
\end{equation}
where $M$ denotes the local order parameter and $M(0)$ is the initial order parameter . This quantity is closely related to the autocorrelation function discussed in the main text. Since $M$ is obtained by averaging spins along a $(d-1)$-dimensional manifold (a line for $2$D), $P(\tau)$ contains an effective integration over a length scale $\xi\sim\tau^{1/z}$, leading to
\begin{equation}
P(\tau)\sim \xi^{d-1}A(\tau).
\end{equation}
In the bulk, using $\xi\sim\tau^{1/z}$ and $A_b(\tau)\sim\tau^{\theta-d/z}$, one has
\begin{equation}
P_b(\tau,L)=\tau^{\theta-1/z}f_7(\tau L^{-z}),
\end{equation}
or equivalently,
\begin{equation}
  P_b(\tau,L)=L^{\theta z-1}\tilde{f}_7(\tau L^{-z}),
  \label{eq:pbsclaing}
\end{equation}
Near the boundary, the two-time correlation is given by
\begin{equation}
P_s(\tau,L)=\tau^{\theta_1-1/z}f_{8}(\tau L^{-z}),
\end{equation}
or equivalently,
\begin{equation}
  P_s(\tau,L)=L^{\theta_1 z-1}\tilde{f}_8(\tau L^{-z}).
  \label{eq:psscaling}
\end{equation}

\begin{figure}[tbp]
\centering
  \includegraphics[width=0.75\linewidth,clip]{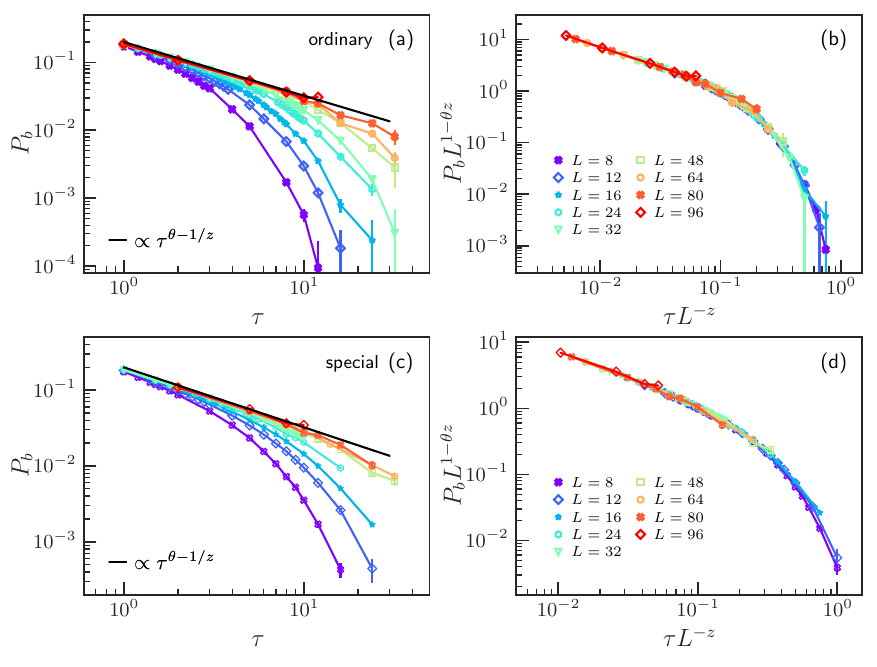}
  \vskip-3mm
  \caption{{\bf Bulk local two-time correlation $P_b$ of the $2$D quantum Ising model.}
    (a) Imaginary-time evolution of $P_b$ for the ordinary transition. The solid line indicates a power-law decay with the exponent $\theta-1/z$.
    (b) Recaling $P_b$ and $\tau$ as $P_bL^{1-\theta z}$ and $\tau L^{-z}$, all data collapse well.
    (c) Similar to (a), but for the special transition.
    (d) Rescaled curves of the data in (c).
  }
  \label{figs3}
\end{figure}

We now verify the scaling form of $P(\tau)$ in the $2$D quantum Ising model. As a benchmark, we consider the bulk first.
Since the bulk critical behavior is governed by the same bulk fixed point for both the ordinary and special transitions, $P_b$ is expected to follow the same asymptotic scaling form in the two cases. As shown in Fig.~\ref{figs3}, for both cases, $P_b\sim \tau^{\theta-1/z}$, with $\theta=0.209(1)$ taking the same bulk critical initial-slip exponent~\cite{Shu2017prb}. After rescaling  $\tau$ and $P_b$ as $\tau L^{-z}$ and $P_bL^{1-\theta z}$, respectively, all data collapse well, confirming Eq.~(\ref{eq:pbsclaing}).

\begin{figure}[tbp]
\centering
  \includegraphics[width=0.75\linewidth,clip]{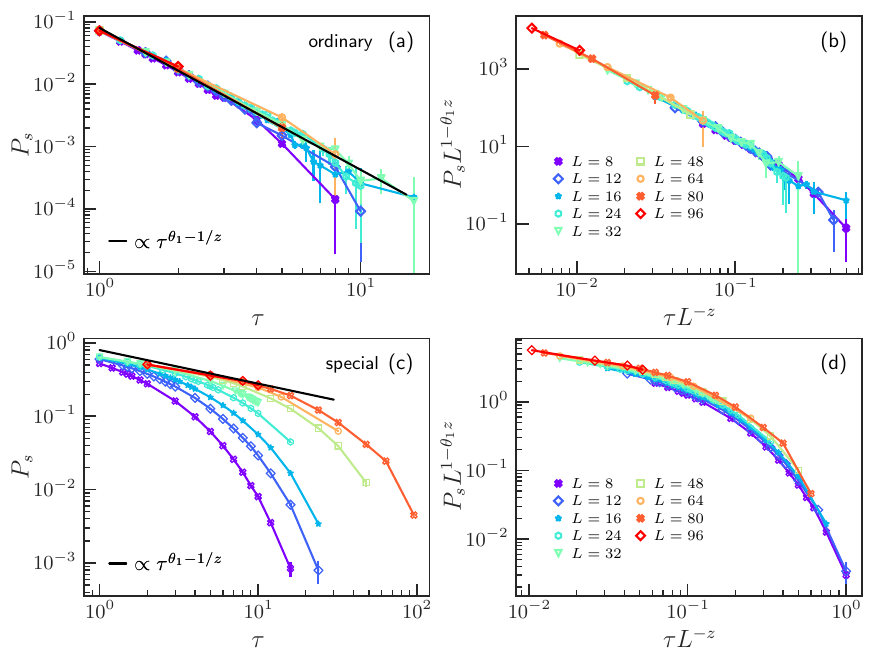}
  \vskip-3mm
  \caption{{\bf Boundary local two-time correlation $P_s$ of the $2$D quantum Ising model.}
    (a) Imaginary-time evolution of $P_s$ for the ordinary transition. The solid line indicates a power-law decay with the exponent $\theta_1-1/z$.
    (b) Recaling $P_s$ and $\tau$ as $P_sL^{(1-\theta_1 z)}$ and $\tau L^{-z}$, all data collapse well.
    (c) Similar to (a), but for the special transition.
    (d) Rescaled curves of the data in (c).
   }
  \label{figs4}
\end{figure}

Near the boundaries, $P_s(\tau)$ exhibits power-law behavior for both the ordinary and special transitions, taking the value of $\theta_1-1/z$ for the corresponding universality classes, as shown in Figs.~\ref{figs4}(a) and (c). The good collapse in rescaled curves in Figs.~\ref{figs4}(b) and (d) provides further support for Eq.~(\ref{eq:psscaling}).

\subsection{Mixed two-time correlation}

We further examine the mixed two-time correlation function
\begin{equation}
  G(\tau)=L^{d-1}\langle M(\tau)m(0)\rangle,
\end{equation}
where $m(0)$ denotes the global order parameter of the initial state. 
Since $G(\tau)$ contains only one local order parameter, the SDE acts only once when $M$ is measured at the boundary.
As a result, $G_s(\tau)$ provides an alternative way to extract $\theta'_1$, circumventing the need to analyze the $m_0\to 0$ limit in the order-parameter relaxation.

For the bulk case, the mixed correlation function scales as
\begin{equation}
G_b(\tau)=b^{-x_m+x_0}G_b(b^{-z}\tau),
\end{equation}
where $b$ is an arbitrary scaling factor, $x_m=\beta/\nu$ and $x_0=x_m+\theta z$. Choosing $b=\tau^{1/z}$, one obtains
\begin{equation}
G_b(\tau)\sim \tau^\theta,
\end{equation}
providing a direct determination of $\theta$.

Near the boundaries, applying the SDE on the local order parameter yields~\cite{Symanzik1981npb,PTCP8,PTCP10,Diehl1997book,Ritschel1995prl,Yin2014prb,Zhang2014pre}
\begin{equation}
M_s(\tau)\sim \tau^{(\beta-\beta_1)/\nu z} M_b(\tau).
\end{equation}
and thus
\begin{equation}
G_s(\tau)\sim \tau^{(\beta-\beta_1)/\nu z} G_b(\tau) \sim \tau^{\theta-(\beta_1-\beta)/\nu z}.
\end{equation}
Therefore,
\begin{equation}
G_s(\tau)\sim\tau^{\theta_1'},
\end{equation}
with
\begin{equation}
\theta_1'=\theta-(\beta_1-\beta)/\nu z.
\end{equation}
Accordingly, the finite-size scaling forms are given by
\begin{equation}
G_b(\tau,L)=L^{\theta z}f_{9}(\tau L^{-z}),
\end{equation}
and
\begin{equation}
G_s(\tau,L)=L^{\theta_1' z}f_{10}(\tau L^{-z}).
\end{equation}

\begin{figure}[tbp]
\centering
  \includegraphics[width=0.75\linewidth,clip]{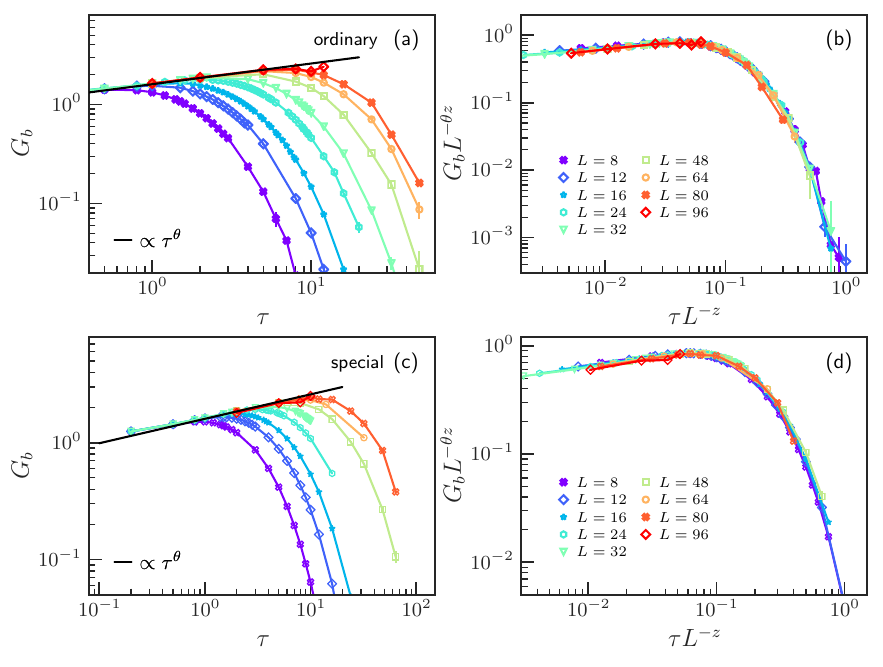}
  \vskip-3mm
  \caption{{\bf Bulk mixed two-time correlation $G_b$ of the $2$D quantum Ising model.}
    (a) Dependence of $G_b$ on $\tau$ for the ordinary transition. The solid line indicates a power-law increase with the bulk critical initial slip exponent $\theta=0.3734$.
    (b) Finite-size rescaled curves of data in (a).
    (c) Similar to (a), but for the special transition, with the same critical initial slip exponent.
    (d) Finite-size rescaled curves of data in (c).}
  \label{figs5}
\end{figure}

Next we verify the above scaling analyses in the $2$D quantum Ising model. Figures~\ref{figs5}(a) and (c) present $G_b(\tau)$ for the ordinary and special transitions, respectively. For both universality classes, the data exhibit a clear power-law increase consistent with the same bulk initial-slip exponent $\theta$.
The initial growth in the short-time regime is captured by $G_b$.
Moreover, after finite-size rescaling, data for different system sizes collapse onto a single curve, supporting the predicted scaling behavior, as shown in Figs.~\ref{figs5}(b) and (d).

\begin{figure}[tbp]
\centering
  \includegraphics[width=0.75\linewidth,clip]{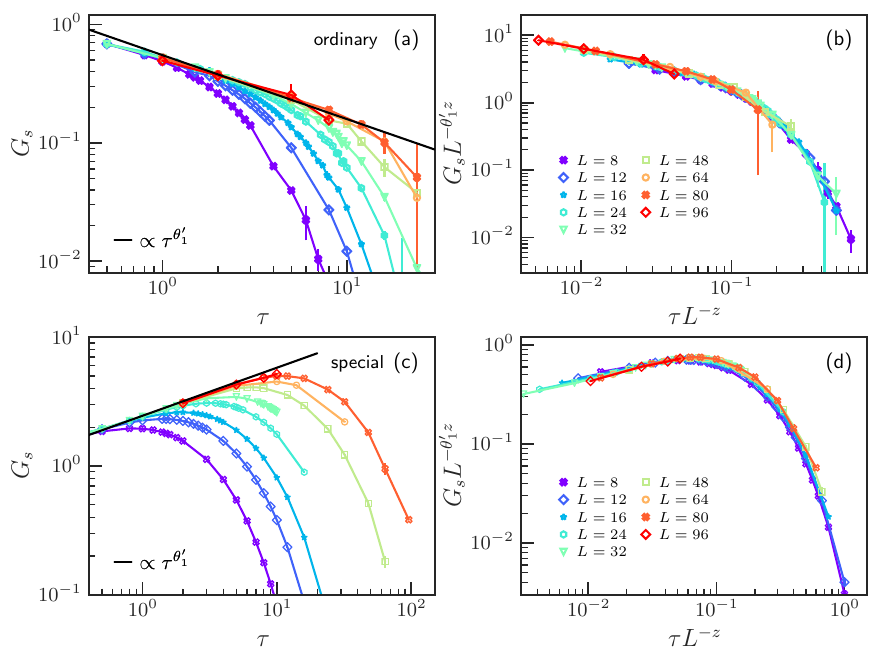}
  \vskip-3mm
    \caption{{\bf Boundary mixed two-time correlation $G_s$ of the $2$D quantum Ising model.}
    (a) Dependence of $G_s$ on $\tau$ for the ordinary transition. The solid line is a power-law decay with the bulk critical initial slip exponent $\theta_1'=-0.532$.
    (b) Finite-size rescaled curves of data in (a).
    (c) Similar to (a), but for the special transition, with the exponent $\theta_1'=0.376$.
    (d) Finite-size rescaled curves of data in (c).
   }
  \label{figs6}
\end{figure}

For the boundary case, the growth of $G_s$ is governed by the boundary exponent $\theta_1'$, reflecting the influence of the boundary on the development of critical correlations. As shown in Fig.~\ref{figs6}(a), $G_s(\tau)$ exhibits a power-law decay with the exponent $\theta'_1=-0.532$ in the ordinary transition, while in Fig.~\ref{figs6}(c), $G_s(\tau)$ increases with the exponent $\theta'_1=0.376$ for the special transition, both agreeing well with the theoretical prediction. The corresponding finite-size scaling collapses further support the scaling form derived above, as shown in Figs.~\ref{figs6}(b) and (d).

In summary, the three correlation functions discussed in this work probe different combinations of bulk and boundary initial-slip exponents. In the bulk,
\begin{equation}
A_b(\tau)\sim\tau^{\theta-d/z},\qquad
P_b(\tau)\sim\tau^{\theta-1/z},\qquad
G_b(\tau)\sim\tau^\theta.
\end{equation}
At the boundary,
\begin{equation}
A_s(\tau)\sim\tau^{\theta_1-d/z},\qquad
P_s(\tau)\sim\tau^{\theta_1-1/z},\qquad
G_s(\tau)\sim\tau^{\theta'_1}.
\end{equation}
The scaling behaviors of $A$, $P$, and $G$ are fully consistent with one another and with the scaling relations derived in the main text, providing a comprehensive verification of the boundary short-imaginary-time scaling theory.

\end{document}